\documentclass[reprint,twocolumn,amsmath,amssymb,superscriptaddress]{revtex4-2}
\usepackage{hyperref}
\usepackage{graphicx}
\usepackage{dcolumn}
\usepackage{bm}
\usepackage{xcolor}
\usepackage{epstopdf}
\usepackage{textgreek}

\usepackage{mathtools}
\usepackage[version=4]{mhchem}
\usepackage{amsmath}
\usepackage{amssymb}
\usepackage{siunitx}
\DeclareSIUnit\angstrom{\protect \text {Å}}

\setcitestyle{super}

\renewcommand*{\overrightarrow}[1]{\vbox{\halign{##\cr 
  \tiny\rightarrowfill\cr\noalign{\nointerlineskip\vskip1pt} 
  $#1\mskip2mu$\cr}}}
\newcommand*{\norme}[1]{\left\lVert\overrightarrow{#1}\right\rVert} 

\begin{document}

\title{
How to steer active colloids up a vertical wall\\
}
\author{Adérito Fins Carreira}
\thanks{These two authors contributed equally}
\affiliation{Universit\'e de Lyon, Universit\'e Claude Bernard Lyon 1, CNRS, Institut Lumi\`ere Mati\`ere, F-69622, Villeurbanne, France}
\author{Adam Wysocki}
\thanks{These two authors contributed equally}
\affiliation{Department of Theoretical Physics and Center for Biophysics, Saarland University, 66123 Saarbr\"{u}cken, Germany}
\author{Christophe Ybert}
\author{Mathieu Leocmach}
\affiliation{Universit\'e de Lyon, Universit\'e Claude Bernard Lyon 1, CNRS, Institut Lumi\`ere Mati\`ere, F-69622, Villeurbanne, France}
\author{Heiko Rieger}
\affiliation{Department of Theoretical Physics and Center for Biophysics, Saarland University, 66123 Saarbr\"{u}cken, Germany}
\affiliation{Leibniz Institute for New Materials INM, Campus D2 2, 66123 Saarbr\"{u}cken, Germany}
\author{C\'ecile Cottin-Bizonne}
\affiliation{Universit\'e de Lyon, Universit\'e Claude Bernard Lyon 1, CNRS, Institut Lumi\`ere Mati\`ere, F-69622, Villeurbanne, France}
\date{\today}

\begin{abstract}

An important challenge in active matter lies in harnessing useful global work from entities that produce work locally, e.g., via self-propulsion. We investigate here the active matter version of a classical capillary rise effect, by considering a non-phase separated sediment of self-propelled 
Janus colloids in contact with a vertical wall. We provide experimental evidence of an unexpected and dynamic adsorption layer at the wall. Additionally, we develop a complementary
numerical model that recapitulates the experimental observations. We show that an adhesive and aligning wall enhances the pre-existing polarity heterogeneity within the bulk, enabling polar active particles to climb up a wall against gravity, effectively powering a global flux. 
Such steady-state flux has no equivalent in a passive wetting layer.\\

\end{abstract}

\maketitle

Self-propelled agents, like bacteria, cells, micro-organisms, animals, or synthetic active particles, inject energy at small scale into their environment, driving the system out of equilibrium.  Often, this energy only powers disordered agitation akin to thermal energy~\cite{tailleurSedimentationTrappingRectification2009} and a major challenge is to better understand active energy flows to extract useful work from seemingly low-grade heat.
This was achieved in a few configurations with, for instance, swimming micro-organisms  driving unidirectional rotation~\cite{diLeonardo2010}, or decreasing the apparent viscosity of their sheared medium~\cite{Rafai2010}, both effects that have no equivalent in equilibrium systems. 
To date, such active energy harvesting did not show up in the ubiquitous configuration of self-propelled particles exposed to a constant and uniform force, e.g. gravity.
There, indeed, active sedimentation properties were shown to be essentially that of a hotter suspension~\cite{palacciSedimentationEffectiveTemperature2010}.
In this paper, we explore how this picture may change drastically when bringing a lateral confining wall in contact with sedimenting polar active particles, a configuration which may fall within active wetting phenomena.

A straightforward mean for harvesting active self-propulsion forces is to herd the system. This can be obtained with a polarizing external field, where particles align with the field and are harnessed to produce useful work. For instance, the swimming direction of magnetotactic bacteria can be polarized by a magnetic field, producing a net flux~\cite{waisbord2016DestabilizationFlowFocused}. Bottom-heavy algae~\cite{kessler1985HydrodynamicFocusingMotile}, colloids~\cite{wolff2013SedimentationPolarOrder, brosseau2021MetallicMicroswimmersDriven, carrasco-fadanelli2023SedimentationLevitationCatalytic} or walkers~\cite{khan2011WetGranularWalkers} may point up, effectively acting against gravity. 
Yet, in the absence of external torque acting on particles, a constant and uniform force field can still promote a local averaged polarization \cite{Enculescu2011, ginotSedimentationSelfpropelledJanus2018}. However, this polarization only arises self-consistently to balance the sedimenting flux such that no net flow is created at steady-state.

Confining walls are known to promote complex responses of active systems which challenge the intuition one has for the equilibrium. At odds with at-equilibrium thermal motion, the pressure exerted by self-propelled particles on walls depends on the wall stiffness~\cite{Solon2015} and curvature~\cite{nikola2016ActiveParticlesSoft}. 
Alike, because of directional persistence~\cite{elgeti2013epl,Schaar2015}, self-propelled particles tend to accumulate at and to polarize towards walls, although the wall-particle interaction, in particular aligning interactions, have a major influence on the detention time of the particles~\cite{Schaar2015}.  

Considering a heterogeneous active system displaying a fluid-fluid interface, the introduction of a wall extends wetting phenomena to active systems. With simple molecular liquids, surface tension that stems from the attractive interactions of the molecules with the wall and between each other~\cite{deGennes2003} can trigger ``wall-climbing'': from the formation of a wetting layer, to a meniscus, to capillary rise. Yet, the steady-state of this equilibrium system displays no net flux. Such phenomenology also extends to complex fluids such as a colloidal suspension with attractive 
interactions~\cite{Aarts2005,hennequin2008FluctuationForcesWettinga}. In this case the interfacial tension between the colloid-rich and the colloid-poor phase is ultralow leading to mesoscopic interface fluctuations. Starting from such equilibrium wetting configuration
of phase separated polymer solutions near a vertical wall, a recent experimental and theoretical study \cite{Adkins2022} demonstrated surprising wetting-like phenomena when adding an active nematic into one of the two phases. In particular, an activity-induced transition towards full wetting was observed.

Even without attractive interactions, active particles can also display a purely out-of-equilibrium separation into a dense and a dilute phase, called motility induced phase separation (MIPS), which is due to a slowdown occurring during collisions between particles~\cite{CatesTailleur2015}. We may thus anticipate for such repulsive active systems a possible extension of the passive scenario, although the mere definition of surface tension is difficult and may yield negative values \cite{Bialk2015, Fausti2021}, which should be incompatible with a stable interface. Indeed, a recent theoretical 
study demonstrates such an analogue out-of-equilibrium behavior, where a repulsive active dense phase can form a meniscus on a wall or rise against gravity in a confining channel due to the slowdown that occurs during particle-wall collisions~\cite{wysockiCapillaryActionScalar2020}.

Here, we present experimental observations of the behavior of a non phase separating gaseous assembly of active colloids that are 
sedimenting in a gravitational field in the presence of a confining wall. We provide the evidence of an unexpected and dynamic adsorption
layer at the wall that has no analogue in passive systems. 
We also offer a numerical model that recapitulates the experimental observations. It enables us to test the respective influences of adhesive and aligning interactions with the wall, parameters that are challenging to adjust experimentally.
Our results demonstrate that a confining wall can act as a pump against a force parallel to it, opening the door to active microfluidic circuits where a configuration as simple as gravity and walls could play a role analog to a generator in an electric circuit.

\section*{Results}

\paragraph*{\bf Experimental and numerical observation of ``Wetting-like'' behaviour --- }

\begin{figure}[tbhp]
\centering

      \includegraphics{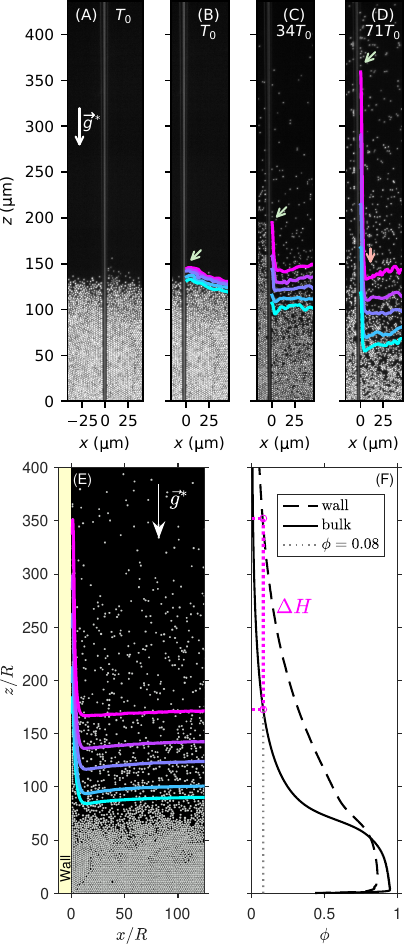}

\caption{{\bf Active Janus colloids climb a wall.} (A) Experimental set-up: glass wall dipped into a sedimented monolayer of passive colloidal particles, at room temperature $T_0$, under a gravity $\Vec{g}^*$. (B-D) Iso-density maps of the colloids at $T_0$ (passive case) and for two activities $34 T_0$ and $71 T_0$. The iso-density values $\phi$ are from top to bottom $0.08$, $0.12$, $0.16$, $0.24$ and $0.3$. The pale green arrow highlights the maximum height of the isodensity curve $\phi=0.08$ at the adsorption layer. The pale red arrow indicates the small depression of the density close to the wall. (E) Snapshot of the
numerical simulation of an assembly of ABPs under gravity near a wall with alignment $\tilde{\Gamma}=6.5$ and adhesion strength $\tilde{\epsilon}=6.5$, respectively, at an effective temperature $64.4 T_0$ (see Methods for details). Isolines at densities $\phi=0.08,\,0.12,\,0.16,\,0.24,\,0.3$ are shown (from top to bottom). (F) Density profiles $\phi(z)$ predicted
by our ABP model far from the wall (solid line), and in the \textit{adsorption layer} (dashed line). The dotted line indicates the density $\phi=0.08$ from which the adsorption layer height $\Delta H$ is defined.}
\label{fig1}
\end{figure}

We study experimentally the behavior of sedimenting active colloidal particles in the presence of a vertical wall. 
The experimental configuration, as shown in Fig.\ref{fig1} and explained in detail in the Methods section, is composed of Janus microspheres of average radius $R=\SI{0.8}{\micro\metre}$, sedimenting along the vertical direction $z$ under an effective gravity $\Vec{g}^*$.
Those particles self-propel through phoretic effects in the presence of hydrogen peroxide \cite{Paxton_jacs-2004, ginotnc2018}. In line with our previous works, we determine the activity of the system by analyzing the sedimentation profile, in the dilute regime, far away from the wall, assuming a Boltzmann distribution. 
From the characteristic sedimentation length, we extract the ratio of the effective to the room temperature, $T_\mathrm{eff}/T_0$, which indicates how far the system is out-of-equilibrium~\cite{Theurkauff_prl-2012, ginotprx15}. 
By changing the concentration of hydrogen peroxide, this ratio can typically be tuned between 1 and 75.

A wall is introduced in the system as a glass capillary oriented along $z$ and immersed down to the colloidal sediment.
To study how this wall impacts the active particles system, we measure the density field $\phi(x,z)$ for various $T_\mathrm{eff}/T_0$.  
In Fig.\ref{fig1} we show images of the sediment together with a few iso-densities (see Methods) in the passive and active cases. As expected for purely repulsive particles with no wall-attraction, passive colloids at $T_0$ (Fig.\ref{fig1} A-B) are hardly affected by the wall \cite{aartsInterfaceDemixedColloid2006} and the passive system does not adsorb at the wall.

As we activate the colloids with hydrogen peroxide ($T_\mathrm{eff}/T_0>1$), iso-densities remain horizontal and parallel to each other far from the wall. This is what is expected for activity resulting in a simply hotter system with a higher sedimentation length (Fig.\ref{fig1} C-D). 
Strikingly, this hot-colloids mapping breaks down in the vicinity of the wall, where we evidence an
upturn of the iso-densities that rise with increasing activity. The wall causes particles to climb up at its contact. Indeed, we observe what seems to be a wetting or an adsorption layer, a layer of one particle thickness.
Past this wall region, we note a small dip of the iso-densities before progressively returning to the far-field ``flat'' region. 
Overall, this departs strongly from the global features of classical passive wetting or of recent extension with active perturbation \cite{deGennes2003, Adkins2022}, where a macroscopic upward meniscus smoothly bridges between the wall and far-field unperturbed region.

In a classical picture, colloids excess at the wall comes from attractive wall-particle interactions. Indeed, as already mentioned in the introduction, wall accumulation of active particles is a generic feature for which directional persistence provides an underlying mechanism for \emph{effective} adhesion \cite{elgeti2013epl, ginotprx15}.
However, in specific systems, 
interactions between walls and active-particles comprise a rich variety of contributions.
For instance a \emph{direct} wall-particle attraction may also arise due to activity through phoretic or hydrodynamic effects \cite{ginotprx15,liebchenInteractionsActiveColloids2022,Ketzetzi2022}.
Likewise, hydrodynamic interaction of dipole micro-swimmers with surfaces can induce alignment with the wall \cite{Das2015, berke2008prl, Liebchen2022, RojasPrez2021}.

To complement our experimental evidences, we also analyzed the
predictions of a model of sedimenting repulsive active Brownian particles (ABPs) \cite{fily2014SM} in the presence of a vertical wall.
As we shall discuss, this allows us to explore the importance of the detailed 
interaction between walls and active-particles. 
When including an activity-induced interaction, mimicking known diffusiophoretic adhesion \cite{ginotprx15} as well as a wall bipolar (also called nematic) alignment parallel to the wall \cite{Das2015, berke2008prl, Simmchen2016} (see Methods for details), the whole experimental phenomenology is reproduced.
As is shown in Fig.\ref{fig1}~E, numerical simulations display both the adsorption 
layer at the wall in the active case and the small dip close to the wall. Note that consistently with experiments, we explored here an activity region for which no MIPS occurs, corresponding to moderate activities
 quantified by $\mathrm{Pe}_s\leq 17$, far enough from $\mathrm{Pe}_\mathrm{crit}\approx 26.7$ 
\cite{siebert2018pre} (see Methods for the definition of the normalized swim persistence length $\mathrm{Pe}_s$).
In the MIPS regime self-propelled particles generate a
phenomenology that is reminiscent of classical wetting configuration
as a recent theoretical study predicted a smooth macroscopic wetting meniscus \cite{wysockiCapillaryActionScalar2020}, in contrast with the present observations.

\paragraph*{\bf Detention time at a wall without gravity --- }
Before interpreting our results, we have to understand how active particles accumulate at a wall in the absence of gravity and the influence of wall-particle interactions on this accumulation. To do this, we generalize arguments that have been laid out in the case of steric \cite{elgeti2013epl} and hydrodynamic interactions \cite{Schaar2015}.
To simplify the argumentation, we neglect the translational diffusion, i.e. $\mathrm{Pe}_s \gg 1$.
In the absence of attractive interactions, an active particle is able to escape the wall as soon as the normal component of its propulsion force points away from the wall. Since its orientation evolves through rotational diffusion, a particle polarized towards the wall will have a long detention time, whereas a particle polarized tangentially to the wall have a shorter detention time.
Therefore, wall-aligning interactions actually reduce detention time because they bring the particles closer to the escape angle.
By contrast, attractive interactions push the escape angle away from the wall and thus increase detention time.
A combination of attractive and wall-aligning interactions traps the orientation parallel to the wall while pushing the escape angle away, and thus increases detention time.
Let us now explore how a force 
parallel to the wall affects these behaviours and how it explains our observation of an activity-induced adsorption layer.
To do this, we use our ABP model to explore systematically the influence of the above-mentioned contributions, namely self-propulsion, direct wall adhesion, wall alignment and the combination of these.

\paragraph*{\bf  Adsorption layer height dependency --- }
From the density field we obtain the density profiles $\phi_{\mathrm{wall}}(z)$ close to the wall (within the adsorption layer) and $\phi_{\mathrm{bulk}}(z)$ in the bulk, far away from the wall (see Methods). 
It is clear from Fig.\ref{fig1}~F that at a given altitude, the adsorption layer has an excess density as compared to the bulk and that the altitude corresponding to a certain density is much higher in the adsorption layer than far from the wall. The height difference between the adsorption layer and the bulk gets even larger as the density decreases. We now define the  
\textit{adsorption layer height}
$\Delta H$ as the difference in altitudes corresponding to a given density of $\phi=0.08$ in the adsorption layer and the bulk. 
$\Delta H$ is a global observable of the activity-induced phenomenon.

\begin{figure*}[tbph]
\centering
\includegraphics[width=1.5\columnwidth]{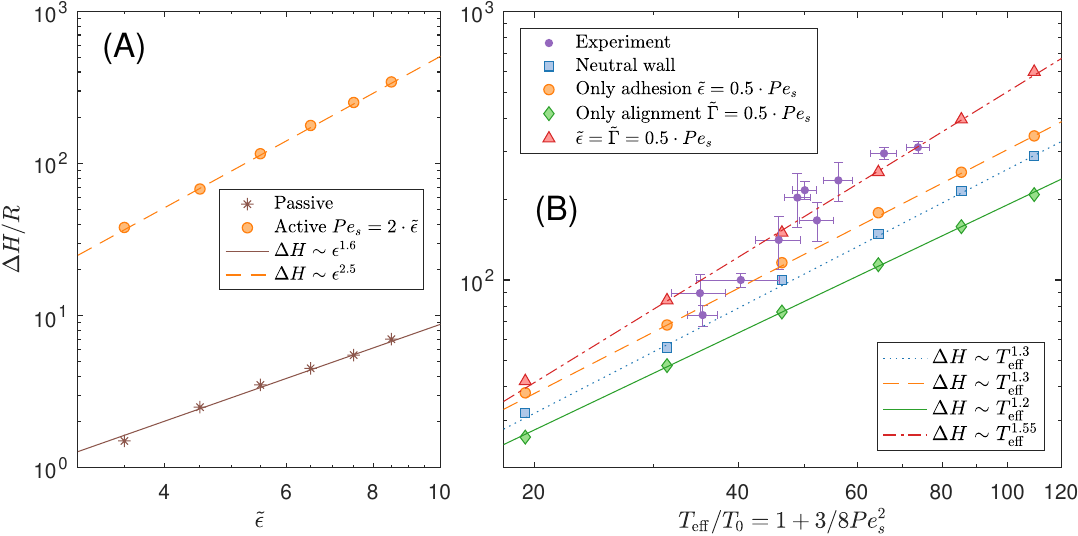}
\caption{%
{\bf Adsorption layer height dependency.} 
(A) Numerical values of the adsorption layer height $\Delta H$ as a function of the wall adhesion strength $\tilde{\epsilon}$ in the passive (brown stars) and active (orange dots) case, where in the latter case we assume that the adhesion strength increases linearly with activity as $\tilde{\epsilon}=0.5 \mathrm{Pe}_s$.  
(B) Adsorption layer height 
$\Delta H$ as a function of the effective temperature $T_{\mathrm{eff}}/T_0$ from experiment (purple dots) and from simulations, for, a neutral wall (blue square), a purely adhesive wall (orange circle) with $\tilde{\epsilon}=0.5 \mathrm{Pe}_s$, a purely aligning wall (green diamond) with $\tilde{\Gamma}=0.5 \mathrm{Pe}_s$ and a wall with both alignment and adhesion (red triangles) with $\tilde{\epsilon}=\tilde{\Gamma}=0.5 \mathrm{Pe}_s$.
}
\label{fig2}
\end{figure*}

For a passive system, wall adhesion promotes the creation of a gravity-fighting adsorption layer \cite{Aarts2005,aartsInterfaceDemixedColloid2006,dimitrov2007prl}, as can be seen in Fig.\ref{fig2}-A where however $\Delta H$ only reaches a few particle radii $R$.
Now, let us consider an active system experiencing both wall accumulation due to the persistence of motion and an activity-induced \emph{direct} adhesion.
Indeed, for Janus colloids, the self-generated electro-chemical gradients responsible for self-propulsion also generate a wall-attraction force, thus scaling as the propulsion velocity (see Supp. Mat. for  an explanation for the form of the adhesive energy, along with simple estimation  on its magnitude). 
Accordingly, we take a wall adhesion strength as $\tilde{\epsilon}\propto \mathrm{Pe}_s$ in our ABP model.
Comparing passive and active systems with same direct adhesion parameter in Fig.\ref{fig2}-A, we observe that $\Delta H$ of self-propelled particles exceeds by more than a decade the one of passive colloids.
This is a further indication that this active system does not merely behave as an equilibrium system regarding adsorption properties, with self-propulsion of individual particles a key factor of the global response.

In Fig.\ref{fig2}-B, we thus focus on self-propelled particles showing $\Delta H$ obtained numerically for different levels of activity as quantified by effective temperatures $T_{\mathrm{eff}}/T_0=3/8\mathrm{Pe}_s^2+1$ \cite{solon2015epjst}. A neutral wall represents the benchmark situation where no direct interaction ---neither attraction nor orientation--- is present aside the short-range steric repulsion.
In line with above discussion, adding a \emph{direct} activity-dependent wall adhesion increases the adsorption layer height as compared to neutral wall.
To mimic the tangential alignment of Janus colloids with a wall, we we choose an alignment strength as $\tilde{\Gamma}\propto\mathrm{Pe}_s$. This is consistent with the experimental characterization of Janus colloids as force dipole microswimmers \cite{Campbel2019} which strengh is proportional to the propulsion speed \cite{berke2008prl} (see Supp. Mat. for an explanation of the form of the alignment strength, along with simple estimation on its magnitude).
Our model predicts that the influence of wall-particle interactions on the 
adsorption layer height is consistent with its influence on detention time without gravity: pure alignment interaction decreases $\Delta H$ as compared to the neutral case, and the combined effect of both alignment and \emph{direct} adhesion increases $\Delta H$ above all other cases considered.
However, we shall see in the following that a wall parallel to gravity also exerts a singular influence on polarity and fluxes.
As a final note on the global 
adsorption layer height $\Delta H$, let us stress that Fig.\ref{fig2}-B also incorporates experimental measurements. As we see, our ABP model recapitulates both the experimental order of magnitude and the dependence on activity.

\begin{figure*}[tpbh]
\centering
\includegraphics[width=1.75\columnwidth]{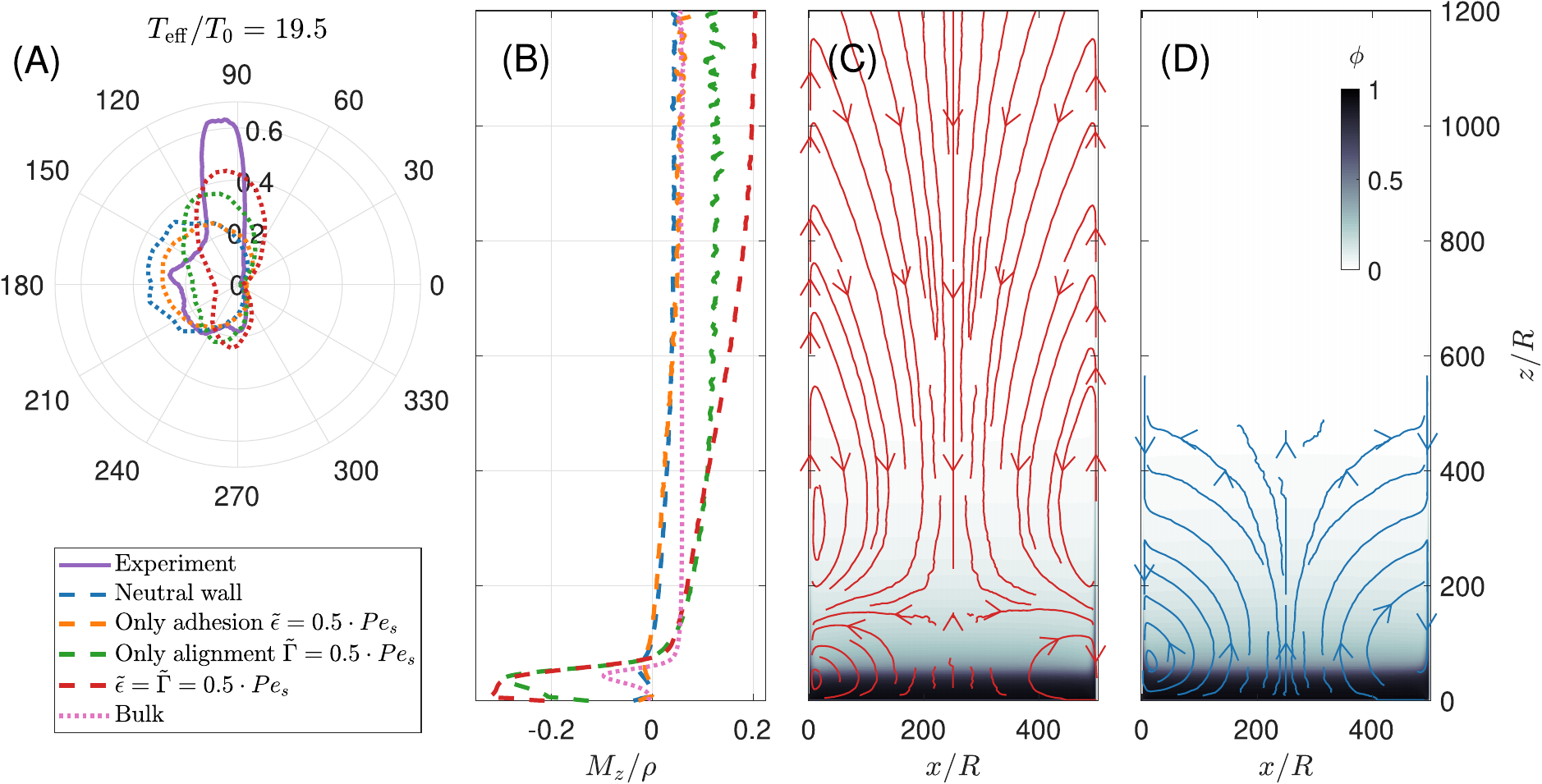}
\caption{%
{\bf Orientation distribution, polarization and fluxes.} (A) Orientation distribution of particles near the wall for $T_{\mathrm{eff}}=19.5 T_0$. We consider in our ABP model (dotted lines) as well as in experiment (purple solid line) the situation which corresponds to the left wall ($x=0$) in Fig 1(E). Thus $\ang{0}$ corresponds to a particle oriented away from the wall and $\ang{90}$ is equivalent to a particle oriented against gravity. (B) The numerical profiles of the $z$-component of the polarization (particle orientation) $M_z/\rho$ in the \textit{adsorption layer} (dashed lines) and in the bulk (dotted line) at $T_{\mathrm{eff}}=109 T_0$ for the four cases: a neutral wall (blue), an adhesive wall (orange), an aligning wall (green) and a wall with alignment plus attraction (red). Particles point upwards if $M_z>0$ and downwards if $M_z<0$. (C,D) Numerical streamlines of a velocity field $\mathbf{v}=\mathbf{J}/\rho$ together with the packing fraction field $\phi$ for a simultaneously attracting and aligning wall with $\tilde{\epsilon}=\tilde{\Gamma}=8.5$ (C) and for a neutral wall (D).}
\label{fig3}
\end{figure*}

\paragraph*{\bf  Polarization at the wall --- } 
To go beyond the global adsorption layer height, we measure numerically the local time- and ensemble-averaged polarization $\mathbf{M}(\mathbf{r})$ (see Methods) and in particular its component parallel to gravity $M_z$ (Fig.\ref{fig3}-B). 
As we already mentioned, a remarkable feature of sedimenting active particles is the existence, in the absence of lateral walls, of a non-vanishing local polarization $M_z(z)$ \cite{Enculescu2011, ginotnc2018}.
In the dilute regime, the mean orientation points upward ($M_z>0$) to balance the downward sedimentation and to
guarantee the absence of a net particle flux. 
By contrast, in the dense sediment at the bottom of the cell there is a downward polarization ($M_z<0$), as the total polarization has to vanish \cite{Hermann2020}. 

Far away from the wall, this bulk behaviour is well recovered in our system (Fig.\ref{fig3}-B) with a local polarization that remains unchanged irrespective of the wall properties details.
At the wall, the same qualitative picture is retained for the averaged polarization, which reverses from an upward to a downward orientation when going from large $z$ down into the sediment.
There, however, the polarization strength becomes very dependent on the wall-particle interaction properties.
For a neutral or a purely adhesive wall, the polarization at the wall is lower than in the bulk. 
Indeed, as recalled above, in the absence of gravity, 
a particle pointing towards the wall will in average remain longer at the wall than a particle nearly parallel to it. 
Mean polarization at the wall is thus obtained from an initial bulk distribution by adding extra weight to normal orientation at the expense of tangential ones, so that vertical non-aligning walls will act as dampers for the bulk $z$-polarization.

On the opposite, a vertical bipolar aligning wall will boost any initial upward or downward polarization bias in the particles orientation. 
This is what we observe (Fig.\ref{fig3}-B) for all aligning configurations (with or without additional adhesion) where the wall acts as an enhancer of the bulk polarization.
In a side experiment, we probed the Janus colloidal particles polarization at the wall (Fig.\ref{fig3}-A). 
The method, fully detailed in Supp. Mat., is based on the tiny shifts between a particle location in the three color channels of a camera that correspond to the colour difference between the two faces of the Janus. As chromatic aberrations are benchmark in the bulk region, only excess polarization at the wall is reported.
We observe a strong upward excess polarization, which is a clear and independent indication of alignment interaction in the experiment.

\paragraph*{\bf  Fluxes, circulations, and wall dynamics ---} 
As we just pointed out, the colloid polarization at a vertical wall is an important feature of the present system.  
With self-propelled objects, this naturally raises the question of permanent fluxes in the system, a property forbidden in equilibrium wetting or adsorption.
Therefore, we calculated
the local particles flux $\mathbf{J(\mathbf{r})}$ (see Methods) predicted by our ABP model and show the result in
Fig.\ref{fig3}-C. We observe an upward flux at the attractive and aligning wall in the dilute regime, consistent with the enhanced upward polarization (Fig.\ref{fig3}-A and B). 
Unlike in the (unconfined) bulk, 
the balance between gravity and upward propulsion is broken at the wall.

This fundamentally out-of-equilibrium response, which involves the presence of permanent fluxes in the steady-state, allows us to propose a rationale for the complex dependence of the rising height $\Delta H$ on aligning wall interactions (Fig. \ref{fig2}-B).
When the alignment comes with an additional \emph{direct} wall adhesion, particles are trapped at the wall  
\cite{Drescher2011,Schaar2015} and display an extra upward polarization in the dilute region as compared to the sedimentation-canceling bulk reference. Accordingly, we expect a high rise of the particles at the wall, which is of purely dynamical origin.
On the contrary, when the wall has only alignment interactions, the extra upward polarization is counteracted by a shorter detention time, which limits the possible rise. Although the final balance is difficult to anticipate, it turns out (Fig. \ref{fig2}-B) to yield a weaker rise than even a simple neutral wall.

Moreover, an aligning wall also enhances the downwards polarization present in the dense part of the bulk. Consistently, it causes a strong downward flux close to the wall, resulting in a well pronounced vortex in the bulk. 
As pointed out, the existence of steady-state particle currents is a strong signature of the non-equilibrium nature of active adsorption \cite{herminghaus2017sm}. Note that, so far, such currents are only accessible in our ABP model, since the velocity measured numerically from the flux is, at maximum, $0.1v_0$, which is lower than the experimental resolution. For the neutral wall (Fig.\ref{fig3}-D), we observe a small downward current close to the wall. Such current can be explained by the polarization decrease at the wall, breaking the nearby flux balance and generating a downward current.

\begin{figure}
\centering
\includegraphics[width=\columnwidth]{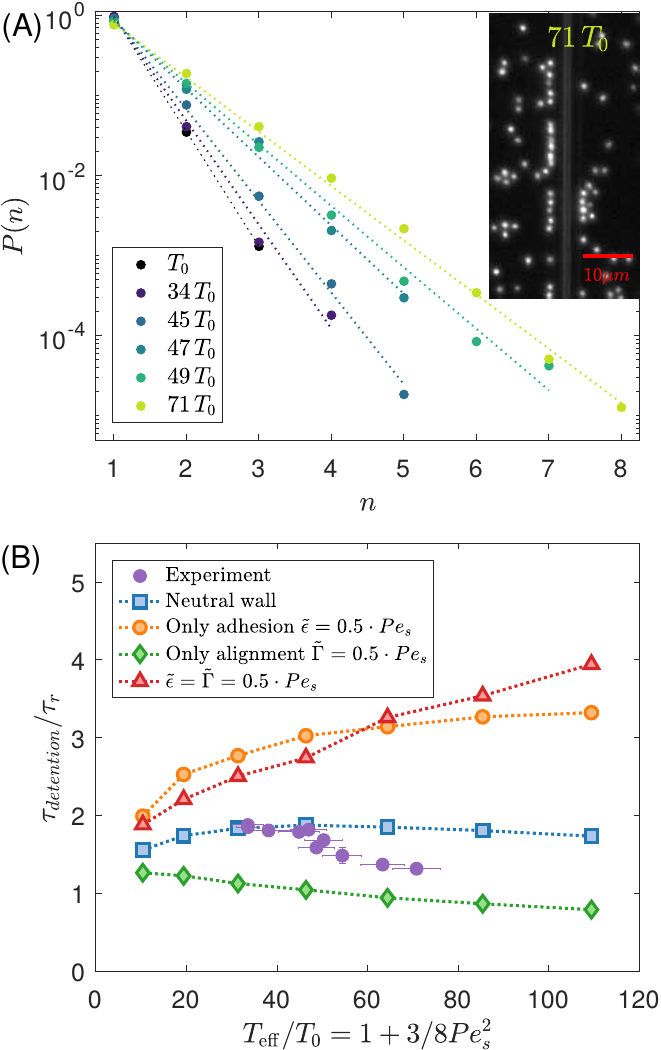}
\caption{%
\textbf{Trains statistics and detention time.} (A) Experimental probabilities of the train size in the adsorption layer $P(n)$ for different effective temperatures. Two colloids form a pair if the distance between the centers is below $\SI{1.9}{\micro m}\approx 2.4R$. 
Dotted lines represent a geometric law fit to the data (See Methods). Inset: Snapshot of active Janus colloids sticking to the wall and forming trains on both sides of the wall. (B) Detention time of colloids without a neighbour in the \textit{adsorption layer} versus the effective temperature $T_{\mathrm{eff}}$. Purple dots indicate the experimental results. We examined different walls in simulations: a neutral wall (blue squares), an adhesive wall (orange circles), an aligning wall (green diamond) and a wall with simultaneous adhesion and alignment (red triangle). 
}
\label{fig4}
\end{figure}

As we just showed, dynamical aspects are the key in the out-of-equilibrium active phenomenon
we report. So far, we have mostly discussed it 
in terms of single particle properties at the wall. However, when zooming in on this wall layer, we see it forms an assembly of 1D clusters, which we denote as \textit{trains}, since they move
collectively along the wall (see inset of Fig.\ref{fig4}-A and movie in Sup. Mat.). 
In a recent study of Janus particles orbiting along a circular post in quasi-2D \cite{Ketzetzi2022}, similar train structures were observed 
resulting from collisions as well as hydrodynamic and osmotic interactions of 1D-moving swimmers. 
Although large trains tend to move slower than single particles at the wall, our trains are dynamical structures. They can merge or break apart. Particles leave the trains at their ends or are squeezed out into the second, more labile, layer. 
We focus in Fig.\ref{fig4}-A on the probability distribution $P(n)$ of train size $n$ in the experiments. We observe that the distribution decays exponentially and thus is completely dominated by monomers.
To describe such behavior, we examine the possible origin of trains in our system, and the extent to which trains can appear randomly, without an underlying formation mechanism. To do so, we consider a simple site-adsorption model where the 
adsorption layer is considered as a 1D system with homogeneously distributed adsorption sites. These sites are randomly populated by adsorption-desorption exchanges with the nearby bulk reservoir (see Methods). The probability of having a train of size $n$ is very well described by such a random site-adsoprtion model, as shown in Fig.\ref{fig4}-A. The dotted lines correspond to geometric laws, where the average density is the only adjustable parameter. The values obtained fall within the range of experimentally measured densities (see Sup. Mat.) 
Overall, the is no indication of collective dynamics on the wall layer which justifies the single particle arguments used so far, although we cannot rule out more subtle collective effects.


Finally, a complete description of the active wall-climbing adsorption 
phenomenon requires a characterization of exchanges between particles in the bulk and in the adsorption layer.
We now report the detention time $\tau_\mathrm{detention}$ that is the average time a particle remains in the adsorption layer
(Fig.\ref{fig4}-B). As monomers dominate train statistics, we focus on the detention time of a single particle. Our ABP model predicts how $\tau_\mathrm{detention}$ depends on the particle-wall interaction. As expected, adhesion always increases detention time as compared to a neutral wall. On the other hand, particles escape a purely aligning wall sooner with increasing activity, however, the opposite is true for a wall that is both adhesive and aligning. These observations are consistent with the wall detention time arguments of \cite{Schaar2015} at $\mathrm{Pe}_s\gg 1$, and with the adsorption layer height dependency on wall-particle interactions, and thus validate our single-particle arguments above.

As shown also in (Fig.\ref{fig4}-B), we were also able to measure $\tau_\mathrm{detention}$ in experiments. There,
we observe $\tau_\mathrm{detention}$ to decrease with activity, which is the same trend as for a purely aligning wall, however, with longer times overall. This 
is consistent with the presence of an alignment interaction, and compatible with an adhesion, however weaker than assumed in our ABP model. 
Nevertheless, for the detention time, there is not a perfect quantitative agreement between the experiments and simulations. Achieving a more accurate agreement would require fine-tuning of the adhesion and the alignment strengths, but beyond that, it would also be necessary to refine the ABP model, particularly by including hydrodynamic and diffusiophoretic interactions \cite{MartnGmez2019, Yan2016, RojasPrez2021, Liebchen2022}.

\section*{Discussion}

To summarize, we study experimentally and theoretically the behavior of an assembly of active colloidal particles in the presence of a vertical confining wall and gravity. We observe a dynamic adsorption layer at the wall that rises with activity, and we find that the interaction between the particles and the wall has a significant impact on this layer. A combination of effective adhesion, alignment  and 
 polarization enhancement at the wall of an already existing
bulk polarization are most likely responsible for the significant increase of density at the wall and the persistent vertical pumping we observe in the system.

How does this persistent pumping fit into the broader context of active matter studies?

The exploration of mechanical aspects in active matter, such as the concept of pressure has triggered an abundant literature illustrating how active matter can depart from equilibrium systems \cite{nikola2016ActiveParticlesSoft, Solon2015}.
These peculiarities of active systems become evident when they interact with walls or interfaces,
at the core of striking features such as work extraction from a bath with ratchet-like rotors~\cite{diLeonardo2010}.
Following this appealing route, the extension of wetting phenomena to active matter was recently considered \cite{wysockiCapillaryActionScalar2020}. In MIPS systems, activity provides an effective attractive inter-particles interaction responsible for the phase separation into a dense active-liquid phase. 
It also provides an effective wall-particle attraction so that phase separated systems in contact with a solid wall display the global phenomenology obeyed by wettable walls in contact with a liquid: macroscopic ascending meniscus, capillary rise, etc. \cite{wysockiCapillaryActionScalar2020}.
Also starting from phase separated systems, but here made of classical polymer-polymer solutions, a rich panel of phenomena were evidenced when adding activity into one of the phases \cite{Adkins2022}. Indeed, activity promoted the transition from a highly wetting meniscus to a full-wetting configuration reminiscent of the Landau-Levich coating transition  \cite{Maleki2011} ---although this was not discussed along this line in \cite{Adkins2022}.
On the contrary, the present system does not phase separate and was only shown to form clusters \cite{Theurkauff_prl-2012, ginotnc2018}. The dense phase is held together mostly because of gravity, and overall no macroscopic meniscus forms at the wall whatever the conditions explored. In that respect, the wall layer that forms with activity would be closer to 
the adsorption of supercritical fluids at solid surfaces \cite{aranovich1996AdsorptionSupercriticalFluids}.
At odds though with this equilibrium analogy, this adsorption layer is associated with strong dynamical effects with steady-state fluxes across the system.

Our results demonstrate that a vertical wall effectively harvests energy from the microscopic scale to produce a macroscopic work. More generally, 
a side wall can act as a pump against a force parallel to it, generating a net steady-state flux in the system. These results pave the way for active microfluidic systems, where 
even a basic configuration involving walls and gravity could play a role analogous to a generator in an electric circuit.

%

%
\section*{Methods}

\paragraph*{Experimental set-up.}
 Gold particles of radius $R=\SI{0.8}{\micro m}$  were grafted with  octadecanethiol \cite{liSelfPropelledNanomotorsAutonomously2015} and 
  half-coated with Platinum  to form Janus microswimmers  when immersed in hydrogen peroxyde (\ce{H2O2}) 
\cite{Paxton_jacs-2004,
 ginotnc2018}. 
Due to their high mass density $\mu\simeq 11\,$g/cm$^3$, 
the particles immediately sediment onto the flat bottom of the experimental cell, forming a bidimensional monolayer of sedimented active particles.
%
A very low in-plane apparent gravity $\vec{g}^*$ is obtained by tilting the whole set-up with a small angle $\theta \approx \ang{0.1}$ in the $z$ direction. 
An elongated borosilicate micropipette bent on the bottom of the observation cell and dipped into the 2D sediment acts as a lateral wall. 
We focus on the half-space on the right side of the wall.

By tuning H$_2$O$_2$ concentration from $c_0= \SI{3.0e{-4}}{v/v}$ up to $5 c_0$,
it is possible to vary the activity of the colloid. 
In practice, for each experiment, we characterize this activity by measuring, from the sedimentation profile in the dilute regime, the ratio $T_\mathrm{eff}/T_0$, where $T_\mathrm{eff}$ and $T_0$ are the effective and the room temperature, respectively \cite{Theurkauff_prl-2012, ginotprx15}.
For each concentration in H$_2$O$_2$, 3000 images  of \SI{2048}{pixel}x\SI{2048}{pixel} are recorded at \SI{5}{fps}  using a Basler camera (ac-A2040-90um) mounted on a Leica DMI 4000B microscope and a custom-made external dark field and a 20x fluotar objective. The pixel size is
$\SI{0.273}{\micro m}$. We track the positions of the center of the particles using Trackpy toolkit on Python~\cite{trackpy2016}. We measure, within one pixel (30\% of $R$) the most occupied position $x_m$ close to the glass wall. 
We set the origin of the $x$ axis ($x=0$) at $0.5 R$ before $x_m$. We define the \textit{adsorption layer} as the layer between $x=0$ and $x=L=2 R$. Note that the average equilibrium distance between particles is $2.4 R$.

Our observable for the analysis are the density maps $\rho(x,z)$, that are obtained by averaging over time the number of particles for each pixel divided by its surface \SI{7.45e-2}{\micro m\squared}. The density profiles of the bulk are computed as $\phi_{\mathrm{bulk}}(z)=\frac{\pi R^2}{x_r-x_l}\int_{x_l}^{x_r}\rho(x,z) \mathrm{d}x$ with 
$x_l= 80 R$ 
corresponding to a value far enough from the wall, at least six times the average equilibrium distance between particles, and $x_r= 340 R$, corresponding to the border of the image.
The density profile in the adsorption layer is obtained via $\phi_{\mathrm{wall}}(z)=\frac{\pi R^2}{L}\int_{0}^{L}\rho(x,z) \mathrm{d}x$. 
From both density profiles, adsorption layer heights can be determined by measuring the difference of altitude using a 0.01 broad interval centered on $\phi=0.08$.

We define the \textit{trains} by considering the set of particles in the adsorption layer whose centers are separated by a distance of less than $2.4 R$. 
The train size distribution was analyzed for trains at altitudes higher than, $z=\SI{300}{\micro m}$  corresponding to a position at which the bulk density is approximately $0.08$ for all studied activities.

\textit{Polarity} is measured 
on different experiments, on the same system, recorded in reflection with a Baumer HGX40c color camera, a 60x objective and a 1.6x zoom. The polarity of a particle is given by the shift between its position on the green channel and its position on the blue channel, corrected for chromatic aberration, as explained in Supplementary Methods.

\textit{Random site-adsorption model.}
 We look at the statistics of trains, that is, of segments of continuously populated sites. We note $p$ the probability that a site is occupied by a particle  and $q=1-p$ the probability that a site is empty. $p$ corresponds to the mean lineic fraction of particles along the wall. The probability that a random chosen site belongs to a train of size $n$ is proportional to $np^n(1-p)^2$. From which we derive the probability of having a train of size $n$ as $P(n)=(1-p)p^{n-1}=q(1-q)^{n-1}$ that abides a geometric law.

\paragraph*{Active Brownian Particle (ABP) model.}
We model the self-propelled particles as two-dimensional active Brownian particles swimming with a constant velocity $v_0$ in a container of size $L_x$ and $L_z$ along the $x$- and $z$-direction, respectively. The position $\mathbf{r}_i=(x_i,z_i)$ and the orientation $\mathbf{e}_i=(\cos{\theta}_i,\sin{\theta_i})$ of the $i$-th particle evolve according to the overdamped Langevin equations:  
\begin{eqnarray}
\dot{\mathbf{r}}_i&=&v_0\mathbf{e}_i+\gamma_t^{-1}\mathbf{f}_{i}-v_g\mathbf{e}_z+\sqrt{2D_t}\,\boldsymbol{\eta}_i\label{eq:eom_pos} \label{eqn1}\\
\dot{\theta}_i&=&\gamma_r^{-1}t_{i}^{\mathrm{wall}}+\sqrt{2D_r}\,\xi_i\label{eq:eom_angle}
\end{eqnarray}
for $i=1,\ldots, N$. The Einstein relation for translation and rotation is $\gamma_t=k_BT_0/D_t$ and $\gamma_r=k_BT_0/D_r$, respectively, where $\gamma_t$ and $\gamma_r$ are the friction coefficients, $D_t$ and $D_r$ the diffusion constants and $k_BT_0$ the thermal energy. $\boldsymbol{\eta}_i$, $\xi_i$ are zero-mean unit-variance Gaussian white noises. For a spherical Brownian particle, we have $D_r = 3 D_t/(2R)^2$. Due to the reduced gravity, particles sediment with velocity $v_g$ along the negative $z$-direction $-\mathbf{e}_z$.
The force on $i$-th particle $\mathbf{f}_{i}$ consists of a part due to particle-particle, $\sum_{j\neq i}\mathbf{f}_{ij}$, and particle-wall interaction, $\mathbf{f}_{i}^{\mathrm{wall}}$. The particles interact via a repulsive pair potential $V(r)=\frac{k}{2}(2R-r)^2$ if $r\leq2R$, i.e., the inter-particle distance $r$ is smaller than the particle diameter $2R$, and $V(r)=0$ otherwise \cite{fily2014SM}. The repulsion strength $k$ is chosen such that the particle overlap is $0.01$ of the diameter $2R$ during a head on collision. The force on $i$-th particle due to $j$-th particle reads as $\mathbf{f}_{ij}=\mathbf{f}(\mathbf{r}_i-\mathbf{r}_j)=-\nabla_{\mathbf{r}_i}V(|\mathbf{r}_i-\mathbf{r}_j|)$. To account for a possible wall adhesion, we let the particles interact with walls via a Lennard-Jones potential, which for the left wall at $x=0$ reads as 
\begin{equation}
    V^{\mathrm{wall}}_\text{left}(x) = 4\epsilon\left[ \left(\frac{R}{x}\right)^{12}-\left(\frac{R}{x}\right)^6 \right]\,,
\end{equation}
where $\epsilon$ controls the attraction to the wall, and similarly for the right, bottom and top wall. For a neutral (purely repulsive) wall, we use a Lennard-Jones potential truncated at $x=2^{1/6}R$ and set $\epsilon=k_BT_0/2$. In order to mimic a possible bipolar (or nematic) wall alignment, the particles experience a position-dependent torque, which for the left wall ($x=0$) reads as
\begin{equation}
t^{\mathrm{wall}}_\text{left}=\Gamma\sin{(2\theta)}\left(\frac{R}{x}\right)^{3}\,,
\end{equation}
where $\Gamma$ is the strength of alignment. For $\Gamma>0$ this torque orients the particles parallel to the wall. The form of $t^{\mathrm{wall}}$ is motivated by hydrodynamic interaction of force dipole microswimmers with surfaces \cite{berke2008prl}, as Janus colloids were experimentally characterized as pushers \cite{Campbel2019}. 
Four dimensionless numbers govern the system. A measure of activity is the normalized swim persistence length, $\mathrm{Pe}_s=v_0/(R D_r)$ sometimes also called swim P\'eclet number. The range of activities considered here was $5\leq\mathrm{Pe}_s\leq 17$, which is below the critical point $\mathrm{Pe}_\mathrm{crit}\approx 26.7$ \cite{siebert2018pre}. Similarly, one can define a gravitational P\'eclet number as $\mathrm{Pe}_g=v_g/(R D_r)$, which we kept fixed to the experimental value $\mathrm{Pe}_g=1$. Furthermore, we define a dimensionless adhesion and alignment strength as $\tilde{\epsilon}=\epsilon/k_BT_0$ and $\tilde{\Gamma}=\Gamma/k_BT_0$, respectively. Our simulation box was of size $L_x=500R$ and $L_z=1500R$ and contained $N=14000$ particles. The simulation time was at least $750000/D_r$. For comparison, the observation time in the experiment was approximately $300/D_r$.

\paragraph*{Microscopic fields.} 
The density field at position $\mathbf{r}$ is given by
\begin{equation}
    \rho(\mathbf{r})=\left\langle\sum_{i=1}^N\delta(\mathbf{r}-\mathbf{r}_i)\right\rangle\,,
\end{equation}
where the angles denote a statistical average, $\delta$ the Dirac delta function, and $\mathbf{r}_i$ the position of $i$-th particle. The microscopic fields are coarse-grained using a Gaussian kernel instead of a delta function \cite{goldhirsch2010gm,monaghan2005rpp}. The local polarization is defined as 
\begin{equation}
    \mathbf{M}(\mathbf{r})=\left\langle\sum_{i}\mathbf{e}_i\,\delta(\mathbf{r}-\mathbf{r}_i)\right\rangle\,,
\end{equation}
where $\mathbf{e}_i$ is the orientation of $i$-th particle. The particles flux $\mathbf{J}(\mathbf{r})$ is obtained from the force density balance equation
\begin{equation}
\mathbf{J}=v_0\mathbf{M}+\gamma_t^{-1}\mathbf{F}-v_g\mathbf{e}_z\rho-D_t\nabla\rho\,,
\label{eq:force_balance}
\end{equation}
where $\mathbf{F}$ denotes the internal force density field \cite{heras2019pre}. Each term of Eq. \ref{eq:force_balance} can be measured easily in simulations.

\bibliography{Wettingarticle}

\section*{Acknowledgements}
A.F.C. is supported by PhD scholarship from the doctoral school of Physics and Astrophysics, University of Lyon.
C.C.B. and C.Y. acknowledge support from ANR-BACMAG.
A.W. and H.R. acknowledge support from the German Research Foundation (DFG), project RI 580/15-1.

\section*{Author contributions}
A.F.C. and A.W. contributed equally to this article.
H.R. and C.C.B. designed the original project.
A.F.C. performed experiments and data analysis.
A.W. performed numerical simulations and data analysis.
M.L. performed the polarity analysis. 
C.Y. provided theoretical justification of the wall interaction used in simulations. 
A.W., C.Y., M.L., H.R. and C.C.B guided the research.
All authors interpreted results and contributed to the manuscript.

\section*{Competing interests}
The authors declare no competing interests.


\section*{Supplementary Material}

\subsection*{Measurement of the excess polarity at the wall}

\begin{figure*}
  \includegraphics[width=\textwidth]{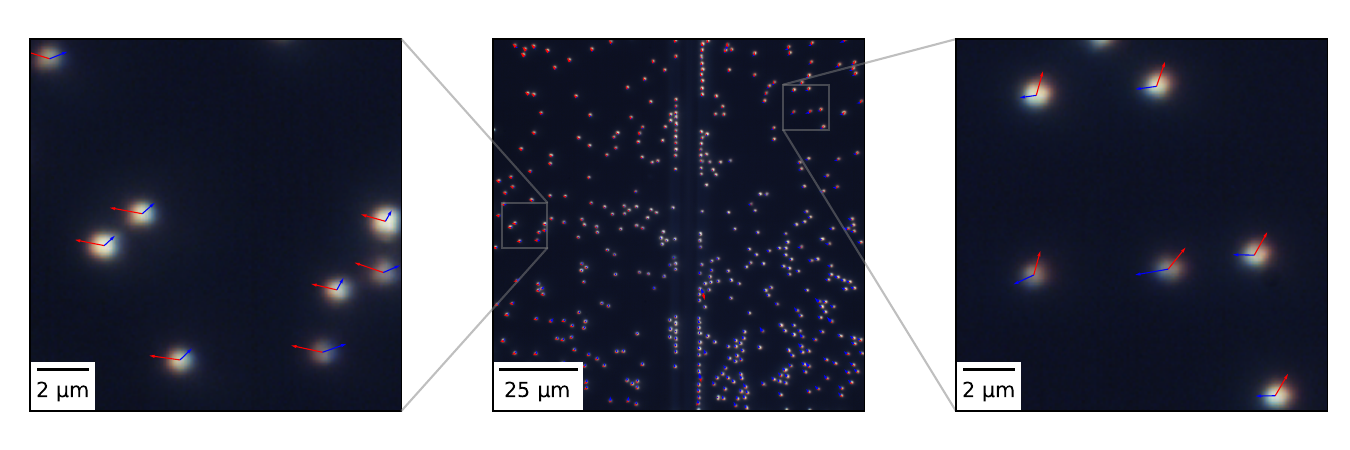}
  \caption{Raw (not corrected for chromatic aberration) redshift and blueshift vectors displayed as red (respectively blue) arrows on top of a white-balanced image. The length of the red and blue arrows are enhanced by a factor 10 for more visibility.}\label{fig:raw_shifts}
\end{figure*}

\subsubsection*{Colour shift}

We acquire RGB images with a Nikon water Immersion x60 objective, NA=1.2 and a x1.6 zoom lens. On an image, a pixel is \SI{57}{\nano\metre} ($\approx 29$ pixels per particle diameter and we have 2048x\SI{2048}{pixel\squared}. Our color camera (Baumer HGX40c) is composed of a monochrome sensor equipped with colour filter following a Bayer matrix. Therefore there are twice as many green pixels as red or blue pixels. Since the green channel is the most spatially resolved, we use it to localize the particle position with subpixel accuray using the package Trackpy~\cite{trackpy2016}. Around the position of each particle, we localise with subpixel accuracy the local maximum of the red (respectively blue) channel. For each particle, we can thus define two vectors: the `redshift' vector (respectively `blueshift' vector) is the difference between the position on the red channel (respectively blue channel) and the position on the green chanel. The results are shown on Fig.~S\ref{fig:raw_shifts}.

On the left zoom, it is obvious that the redshift vectors are mostly oriented towards the left and the blueshift vectors mostly towards the top right. However on the right zoom the redshift vectors are oriented towards the top right and the blueshift vectors are oriented towards the left. This means that we have chromatic aberration in our optical system: the different wavelength of light are not propagated in the same way, therefore the red image does not form at the same place on the camera as the green image or the blue image. The raw redshift and blueshift vectors must be corrected of this chromatic aberration before they can yield any information on particle polarity.

\subsubsection*{Chromatic aberration}

\begin{figure*}\centering
  \includegraphics[width=0.7\textwidth]{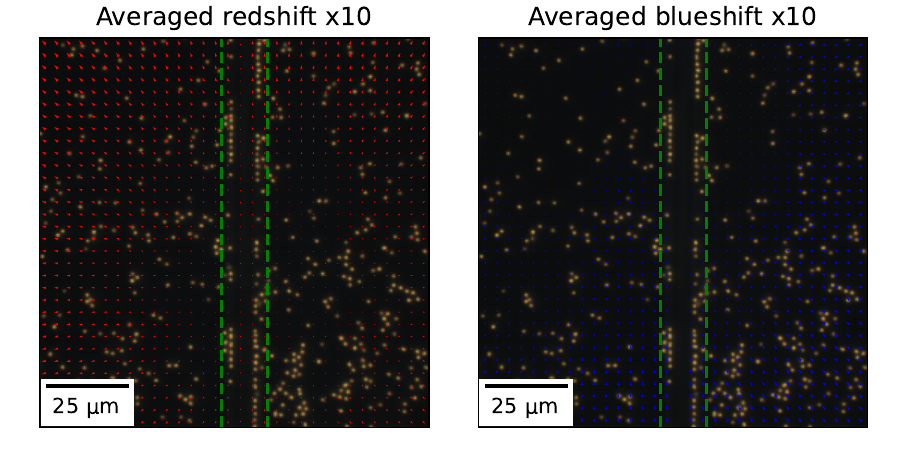}
  \caption{Chromatic aberration observed by averaging redshift and blueshift on a grid and in time. The length of the red and blue arrows are enhanced by a factor 10 for more visibility. The vertical green line delimit the region of the pipette.}\label{fig:average_shifts}
\end{figure*}

In order to quantify and correct for the chromatic abberation, we divide the field of view into a $32\times 32$ grid and average the redshit and blueshift aberration of all the particles that are in a grid element at a given time, for all 6000 images of the video. Since the polarity of chemically bound dimers can be ill defined, we removed all pairs of particles closer than 28px from this ensemble average. Results are shown on Fig.~S\ref{fig:average_shifts}.

We want to remove the global chromatic aberration, but not all polarity signal, especially close to the wall. Therefore, we remove from our grid all cells that can contain particles at walls, i.e. what is between the two vertical green dashed lines on Fig.~S\ref{fig:average_shifts}. The remaining field is interpolated using splines.

\begin{figure*}
  \includegraphics[width=\textwidth]{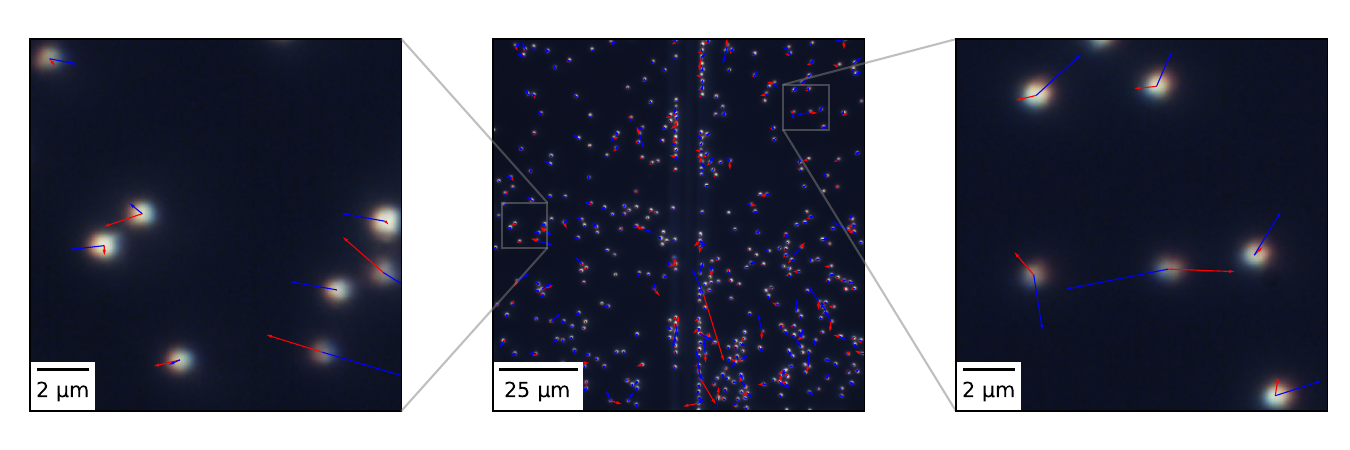}
  \caption{Chromatic aberration-corrected redshift and blueshift vectors displayed as red (respectively blue) arrows on top of a white-balanced image. The length of the red and blue arrows are enhanced by a factor 100 for more visibility.}\label{fig:corrected_shifts}
\end{figure*}

Now, at each instantaneous position of a particle, we can remove from redshift or blueshift the value of the spline at this position. The remaining redshift and blueshit are shown on Fig.~S\ref{fig:corrected_shifts}, this time magnified by a factor 100. The abberation-corrected redshift and blueshift are typically 10 times smaller than the value of the chromatic aberration. Therefore, the arrows are now decorrelated from the apparent color shift on the image.

Many particles have a redshift or blueshit measurement that is below the resolution we can expect from particle localisation from images (0.1 px, here 0.2 px since we subtract two measures). Therefore, we cannot measure precisely the polarity of a given particle, we have to
estimate statistical distributions on large numbers of particles. In order to minimze the noise, all the following statistical distributions will be established by weighting a particle contribution by the magnitude of its aberration corrected redshift and/or blueshift.

Furtheremore, by subtracting the average of the redshift or blueshift far from any wall, we removed not only the chromatic aberration, but also any non zero polarity of the bulk. If corrected redshift or blushift contain polarity information, it is the polarity near the wall
with respect to the bulk polarity, and not the absolute polarity.

In the following we consider only the abberation-corrected redshift and blueshift and thus drop the adjective.

\subsubsection*{Statistical correlation between colour shift and polarity}

\begin{figure*}
  \includegraphics[width=\textwidth]{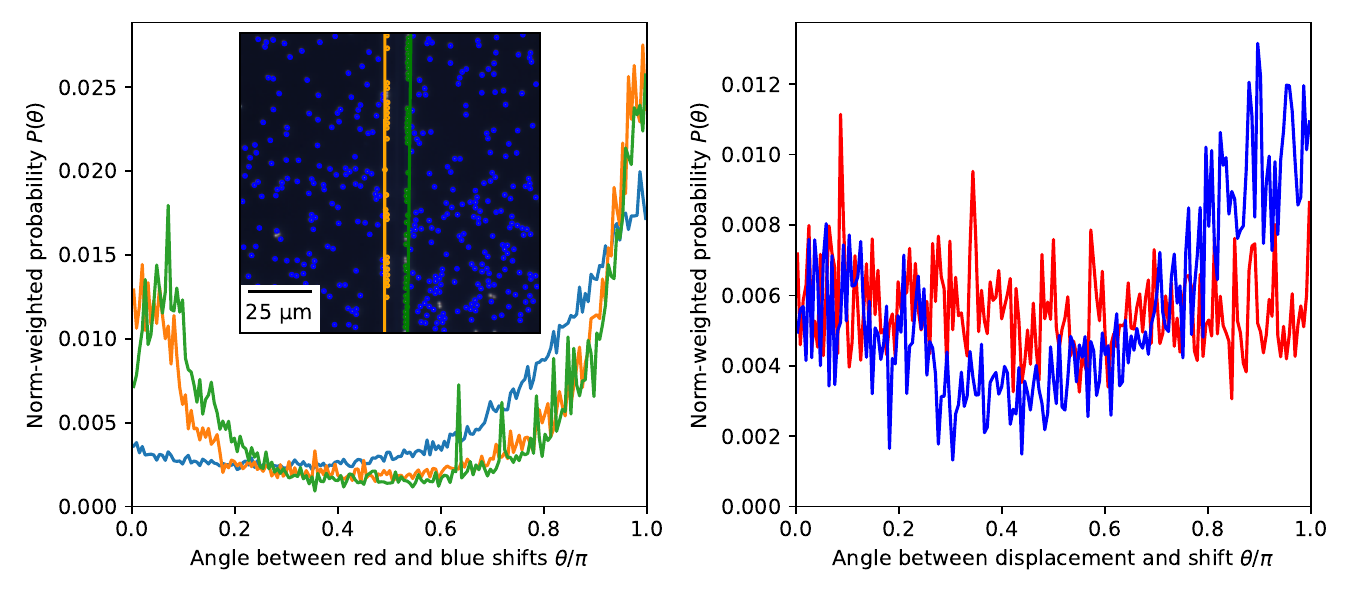}
  \caption{Left: Probability distribution of the angle between abberation-corrected redshift and blueshift of the same particle. Colors correspond to particles in bulk, left wall and right wall, as indicated on the inset.
Right: Probability distribution of the angle between the displacement of a particle between $t$ and $t+\delta t$ and the abberation-corrected redshift (respectively blueshift) of this particle at time $t$ averaged over particles in bulk and time $t$.}\label{fig:angular_distributions}
\end{figure*}

First, we can compute the probability distribution function of the angle $\theta^\text{red blue}_i$ between redshift $\overrightarrow{S^\mathrm{red}_i}$ and blueshift $\overrightarrow{S^\mathrm{blue}_i}$ of particle $i$: 
\begin{equation}
  P^\text{red blue}(\theta) = \frac{\sum_i S^\mathrm{red}_i S^\mathrm{blue}_i \delta(\theta^\text{red blue}_i - \theta)}{\sum_i S^\mathrm{red}_i S^\mathrm{blue}_i},
\end{equation}

where $\delta$ is the binning function, $S^\mathrm{red}_i = \norme{S^\mathrm{red}_i}$ and dimers are excluded from the sums. Results are shown on Fig.~S\ref{fig:angular_distributions} (Left).

Redshift and blushift are statistically antiparallel in bulk, but the peak is quite large. At wall, they are cleany split between two populations, either antiparallel (majority) or parallel (minority).

Then, we compute the probability distribution function of the angle $\theta^\text{red}_i(t)$ between redshift at time $t$ and the displacement $\overrightarrow{\Delta x}(t)$ of the particle $i$ between $t$ and $t+\delta t$, where $\delta t = \SI{100}{\milli\second}$ is the time interval between two successive frames:
\begin{equation}
  P^\text{red}(\theta) = \frac{\sum_t\sum_i S^\mathrm{red}_i \delta(\theta^\text{red}_i - \theta)}{\sum_t\sum_i S^\mathrm{red}_i},
\end{equation}
where sums exclude dimers and particles at the wall. We define $P^\text{blue}(\theta)$ in the same way. Both distributions are shown on Fig.~S\ref{fig:angular_distributions} (Right).

As $P^\text{red}(\theta)$ is flat, the redshift is not correlated to displacement and is thus not a measure of polarity. By contrast, $P^\text{blue}(\theta)$ shows a peak around $\pi$, meaning that blue shift is mostly antiparallel to displacement.

Even far from the walls, particle displacement is an indirect measure of polarity, as particles experience forces other than self-propulsion, i.e. Brownian motion, gravity and particle-particle interactions of electrostatic and hydrodynamic nature. In addition, automatic trajectory reconstruction is subject to tracking errors, which further decrease the signal to noise ratio. Despite these limitations, blueshift is a good \emph{statistical} predictor of particle displacement and polarity.

\subsubsection*{Rationale of blueshift}

\paragraph{Why is blueshift a good measure of polarity and not redshift?}
The platinum cap is thin, thus the light reflected by the particles is mostly yellow, that is to say a mix of red and green with much less blue. The difference between red position and green position (redshift) thus contains little information. By contrast, platinum reflects more
equally the colors, therefore the excess of blue with respect to green (or red) indicates the platinum side.

\paragraph{But why are redshift and blueshift antiparallel?}
The filters on a colour camera slightly overlap. In particular the green channels also captures some red on one side and some blue on the otherside, whereas there is very little cross-talk between red and blue channels. With respect to the red position, the green position is thus slightly shifted towards the blue position. This shift, is enough to cause the statistical correlation between redshift and blueshift, but the signal is so small with respect to noise that the redshift is a poor predictor of polarity and even poorer predictor of the direction of motion.

\subsection*{Wall interaction for Janus self-phoretic particles}
We discuss hereafter the form and magnitude of the wall--particle interaction used in numerical simulations.
\subsubsection*{Diffusiophoretic wall attraction}
Gold-Platinum Janus particles have been shown to experience an activity-induced adhesive interaction of phoretic origin \cite{ginotprx15}. Indeed each catalytic particle act as a chemical monopole source from the far-field creating environmental gradients to which each swimmer responds.
One thus expects that such attractive interaction is experimentally related to the phoretic mechanism responsible for the self-propulsion for which the associated swim force reads
\begin{equation}
	F_\mathrm{swim} = \frac{k_BT_0}{D_t} v_0.
\end{equation}
Setting that the diffusiophoretic adhesion with the wall therefore involves an adhesion energy $\epsilon_\mathrm{phor.}\sim F_\mathrm{swim} R$, we thus expect an adhesion of dimensionless form
\begin{equation}
	\widetilde{\epsilon}_\mathrm{phor.} = \frac{3}{4}\mathrm{Pe}_s,
\end{equation}
where as in the main text $\widetilde{\epsilon}_\mathrm{phor.} = \epsilon_\mathrm{phor.}/k_BT_0$ and $\mathrm{Pe}_s = v_0/(RD_r)$.

Overall, this justifies the form used in numerical simulation for the wall adhesion $\widetilde{\epsilon} = \alpha\mathrm{Pe}_s$, where moreover the constant $\alpha=0.5$ consistent with experiments is very close with the simple guess 3/4.

\subsubsection*{Wall aligning interaction}
As stated in the main text, the aligning torque was taken based on the effect of hydrodynamic interactions between the wall and a pusher swimmer \cite{berke2008prl}. Indeed this was argued to be the dominant contribution for phoretic Janus particles \cite{Das2015} with an associated torque magnitude $t$ reading for small deviations from wall alignment
\begin{equation}
	t = k_BT_0 \left(\frac{R}{R+h}\right)^3 (2\Delta\theta) \times \frac{1}{8}\mathrm{Pe}_s,
\end{equation}

As compared to the form adopted in simulations this amounts to a dimensionless torque magnitude of $\widetilde{\Gamma}_\mathrm{hydro.} = \frac{1}{8}\mathrm{Pe}_s$, again in fair agreement with the magnitude used to compare with experiments $\widetilde{\Gamma}\sim0.5\mathrm{Pe}_s$.

\end{document}